\documentclass{article}

\usepackage{inputenc}
\usepackage{graphicx}
\usepackage{babel}
\usepackage{diagbox}
\usepackage{xcolor}
\usepackage{multicol}
\usepackage{biblatex}
\usepackage[legalpaper, margin=2cm]{geometry}
\usepackage{hyperref}

\begin{document}

\begin{center}

\Large{\bfseries{Optimization of laser pumping for the generation of multi-pulse structures in fiber lasers}}\\
\large
Alix MALFONDET$^{1}$, Guy. MILLOT$^{1,2}$, Philippe GRELU$^{1}$, and Patrice TCHOFO-DINDA$^{1}$\\
$^{1}$Laboratoire Interdisciplinaire Carnot de Bourgogne, UMR 6303 CNRS, Univ. Bourgogne, 9 Av. A. Savary, B.P. 47870, 21078 Dijon Cedex, France\\
$^{2}$Institut Universitaire de France, 1 Rue Descartes 75005 Paris, France
\end{center}
\normalsize

%

%



\begin{abstract}
We address the challenge of configuring a fiber laser cavity to enable efficient access to multi-pulse structures such as dissipative soliton molecules.  We theoretically compare multi-pulsing routes in the parameter space of the laser.  By using a two-dimensional parameter space,  we experimentally demonstrate an important reduction in the laser pumping power required to form a soliton molecules. 
\end{abstract}

\section{Introduction}
\label{Introduction}

In ultrafast fiber lasers,  pulse-to-pulse interactions can give rise to various types of 
multi-pulse structures of great interest,  such as soliton molecules \cite{Tang2001,Grelu2002,Ortac2010,Krupa2017,Liu_2018}, 
soliton crystals \cite{Komarov2008,Andrianov2021}, or molecular complexes \cite{ZWang:18,He:21},
 in which light pulses may execute dynamics similar to those of atoms in condensed matter. 
However, generating these structures on demand remains a major challenge,  due to the general lack of analytical relationship between the laser parameters and the targeted multi-pulse structure. 
Until now, the generation of those solitonic structures requires a tedious adjustment of accessible cavity components, such as the saturable absorber parameters and the pumping power. 
Multi-pulse structures are often generated at high pumping power, leading to relatively high peak power in the cavity, which may exacerbate nonlinear effects and distortion phenomena in the 
pulse profile \cite{Igbonacho2019}. 
 
A recent work has unveiled the theoretical concept of a mode-locked fiber laser (MLFL) that uses a 
tunable band-pass filter (BPF) to generate multi-pulse structures at significantly reduced pump powers,
 while eliminating distortion phenomena in the pulse profile \cite{TchofoDinda2023}.
Given the lack of experimental proof of the concept reported in Ref. \cite{TchofoDinda2023}, the present study makes two major advances: 
the first is to show experimental proof of the feasibility and effectiveness of this theoretical concept.
In addition, we show that by carefully adjusting the central wavelength of the BPF to a value close to the
 effective peak gain of the EDFA (Erbium-Doped Fiber Amplifier) chosen as the gain medium, 
i.e. 1.53$\mu$m (instead of 1.55$\mu$m as suggested in Ref. \cite{TchofoDinda2023}), 
we obtain a much greater reduction in the pumping power for access to the multi-pulse regime of the laser.

\section{Numerical simulations}
\begin{figure}[h]
\centering
\includegraphics[width=.6\linewidth]{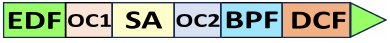}
\caption{Schematic of the fiber laser cavity.}
\label{schema_cavite_lineaire}
\vspace*{-0.2cm}
\end{figure}

For numerical modeling of the ultrafast  laser dynamics, we consider a unidirectional ring cavity with components placed in the order indicated in Fig. \ref{schema_cavite_lineaire} \cite{TchofoDinda2023}.
A section of erbium-doped fiber (EDF) is used as the gain medium. A saturable absorber (SA) is used to trigger mode locking. A BPF with tunable bandwidth is used to optimize the system configuration, and a dispersion compensating fiber (DCF) is used to set the cavity average dispersion to zero.

The action of SA is modeled as follows:
\begin{eqnarray}
P_o = T \,P_{i},\quad  T\equiv T_{0}+\Delta T P_{i}/(P_{i}+P_{sat}),
\end{eqnarray}
where $T$ describes the transmission of the instantaneous SA,
  $T_{0}$  is its transmitivity at low signal, and $\Delta T$  the absorption contrast, while $P_i$  ($P_o$) designates the instantaneous input (output) optical power.
   
		The BPF spectral profile is modeled by the following super-Gaussian function \cite{TchofoDinda2023}:
\begin{eqnarray}
F_{BPF}\left(\omega\right)=exp\left[-2^{2m}\log\left(2\right) \frac{\omega^{2m}}{\Delta\Omega^{2m}_{BPF}}\right],
\end{eqnarray}
where $\Delta\Omega_{BPF}$ is the filter's bandwidth, while $m$ is an integer that determines the filter's profile. The value $m=1$ corresponds to a Gaussian profile, whereas
for $m>$1 the BPF's profile has an increasingly flat top and steeper flanks as m increases.
Throughout the present work, we model the flat-top filter with $m=4$.

\begin{figure}[h]
\centering
\includegraphics[width=.65\linewidth]{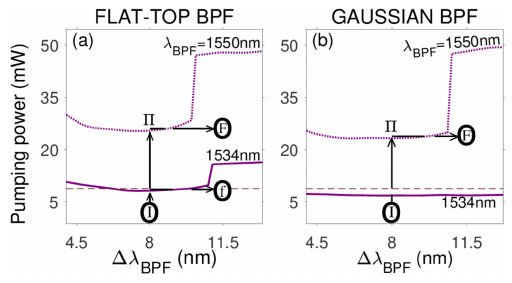}
\caption{Fragmentation pumping power vs filter bandwidth, 
 for a cavity with 1.55$\mu$m-centered BPF (dotted curves) and 1.534$\mu$m-centered BPF (solid-line curves).
 The dashed horizontal line corresponds to the 
FPP for the BPF-free cavity.}
\label{fig_pf_vs_del_ol_2023}
\end{figure}

Two output couplers (OC) are inserted into the cavity.  The coupler OC1 is placed at the output of the amplifier section. The coupler OC2 is placed after the saturable absorber. 
Since all lumped components of the cavity are connected to single-mode fibers, the cavity is therefore equipped with an active fiber and a number of single-mode fiber (SMF) sections.
Using a scalar model, the pulse propagation in these fiber sections can be described by the generalized 
nonlinear Schr{\"o}dinger equation \cite{TchofoDinda2023}.
Numerical simulations are performed using the following parameters:
\begin{center}
\begin{tabular}{|c|c|c|c|}
\hline
\diagbox{\bf{Parameters}}{\bf{Fiber}} & EDF & DCF & SMF \\
\hline
Effective mode area  [$\mu m^2$] & 28.3 & 20 & 78.5 \\
\hline
SOD [ps$\cdot$nm$^{-1}$$\cdot$km$^{-1}$] & -15 & -91.7 & 18 \\
\hline
TOD [ps$\cdot$nm$^{-2}$$\cdot$km$^{-1}$] & $\sim$0 & -0.12 & 0.07 \\
\hline
Total length [m] & 1.2 & 2.9 & 15.81 \\
\hline
Loss [dB/km] & 0.2 & 0.6 & 0.2 \\
\hline
\end{tabular}
\end{center}
where SOD and TOD stand for second-order dispersion and third-order dispersion, respectively.  We set : T$_0$ = 0.7 (70$\%$), $\Delta T$ = 0.3 (30$\%$), P$_{sat}$ = 10 W for the SA. The output coupler transmissions for OC1 and OC2 are set to 90$\%$ and 80$\%$, respectively. The conventional procedure for generating a multi-pulse structure is a two-step process \cite{Li2015,Wang:17,Alsaleh18,zhang:18}. The first is to gradually increase the pumping power (PP) of the laser up to the mode-locking threshold, so as to generate a single pulse in the cavity. In the second step, the PP is progressively raised to the level where the initial pulse splits into several pulses. This value of PP is called fragmentation pumping power (FPP). We applied this conventional procedure to our cavity equipped with the tunable BPF,  by setting its bandwidth $\Delta\lambda_{BPF}$ to different values ranging from 4nm to 13nm. For each value of $\Delta\lambda_{BPF}$, we determined the FPP, and obtained the results shown in 
Fig. \ref{fig_pf_vs_del_ol_2023}. 
The dotted curves in panels (a) and (b) of Fig. \ref{fig_pf_vs_del_ol_2023}, represent the evolution of the FPP as a function of the filter's bandwidth, for a 1.55$\mu$m-centered BPF. 
 We observe that the FPPs are distributed 
over two broad levels, and there exists and a critical bandwidth, 
  $\Delta\lambda_c$, where the FPP jumps from one level to another.  
The PP of the upper level is almost twice as high as that of the lower level, which gives these curves a staircase profile.
Now, we would like to point out that inserting a BPF into a laser cavity not only increases the laser emission threshold, but can significantly increase the FPP.
Indeed, the FPP in the BPF-free cavity is only 8.25 mW, which is represented by the dashed horizontal line in panels \ref{fig_pf_vs_del_ol_2023} (a) and \ref{fig_pf_vs_del_ol_2023} (b), whereas
in the cavity equipped with a flat-top BPF centered at 1.55$\mu$m, the FPP jumps to 26mW for $\Delta\lambda_{BPF} < \Delta\lambda_{c} $, and 49mW for $\Delta\lambda_{BPF} > \Delta\lambda_{c} $.
In this example, by taking realistic BPF transmission and EDF gain features,  the use of this BPF causes an increase of the FPP by a factor of 3 when $\Delta\lambda_{BPF} < \Delta\lambda_{c} $, and by a factor of 6 when $\Delta\lambda_{BPF} > \Delta\lambda_{c} $. Such a high operational cost, added to the cost of this component, clearly raises the question of the practical interest of inserting a 1.55$\mu$m-centered BPF in our laser cavity as suggested in previous work \cite{TchofoDinda2023}.

In the present work, we have resumed numerical simulations with a BPF having a different centering, namely at 1.534$\mu$m.  The results are represented by the solid line curves in panels \ref{fig_pf_vs_del_ol_2023} (a) and \ref{fig_pf_vs_del_ol_2023} (b). They show a dramatic decrease in FPP compared to the case where $\lambda_{BPF} =1.55\mu m$. The most striking point is that for $\lambda_{BPF}$=1.534$\mu$m and $\Delta\lambda_{BPF} < \Delta\lambda_{c} $, the FPP even falls slightly below the level corresponding to the cavity without BPF. 

At first glance, this result may seem surprising because the BPF is a passive component that inevitably increases the level of linear losses in the cavity, and raises the laser emission threshold. In this regard, it should be noted that in our cavity, the laser starts emitting as soon as the PP is high enough for the intra-cavity gain to compensate for the losses, but the wave generated at the emission threshold is a continuous wave, because the nonlinear losses due to the saturable absorber are not yet significant compared to the linear losses. While the insertion of the BPF into our cavity increases the linear losses and the laser emission threshold, the intra-cavity dynamics determine the level of nonlinear losses due to the saturable absorber. It should also be noted that the intra-cavity dynamics are very sensitive to the spectral profile of the BPF.
 Regardless of the type of cavity (with or without BPF), once the laser emission threshold is reached, additional PP is required to access the mode-locking threshold. The amount of this additional PP strongly depends on the intra-cavity dynamics, and more particularly on the spectral profile of the intra-cavity filtering effect. It should be kept in mind that in the BPF-free cavity, a filtering effect (induced by the erbium ion gain curve) remains in the cavity.
Thus, although insertion of the BPF into the cavity increases the level of linear losses in the cavity, its spectral profile may generate intra-cavity dynamics in which the level of nonlinear losses due to the saturable absorber is significantly lower than that generated in the BPF-free cavity.
 In other words, by optimizing the spectral profile of the BPF, the decrease in nonlinear losses can thus counterbalance the increase in linear losses and lead to a FPP of the same level as that of the BPF-free cavity, or even slightly lower.
It can be deduced that, in some bandwidth domains, the BPF centered on 1.534$\mu$m generates an intra-cavity dynamic that strongly favors access to the multi-pulse regime, compared to the dynamics induced by the BPF gain curve, which allows to overcompensate the linear losses induced by the BPF in the cavity.


Thus, our simulations reveal that the best configuration corresponds to a Gaussian BPF centered at 1.534$\mu$m, as it leads to a FPP that is even slightly lower than in the cavity without BPF, as shown by the solid line curve of panel \ref{fig_pf_vs_del_ol_2023} (b).  
In addition, with this type of filter, the transition point disappears from the range of bandwidth values considered in our study, i.e., $4nm \leq \Delta\lambda_{BPF} \leq 13 nm$.

\begin{figure}[h]
\centering
\includegraphics[width=.55\linewidth]{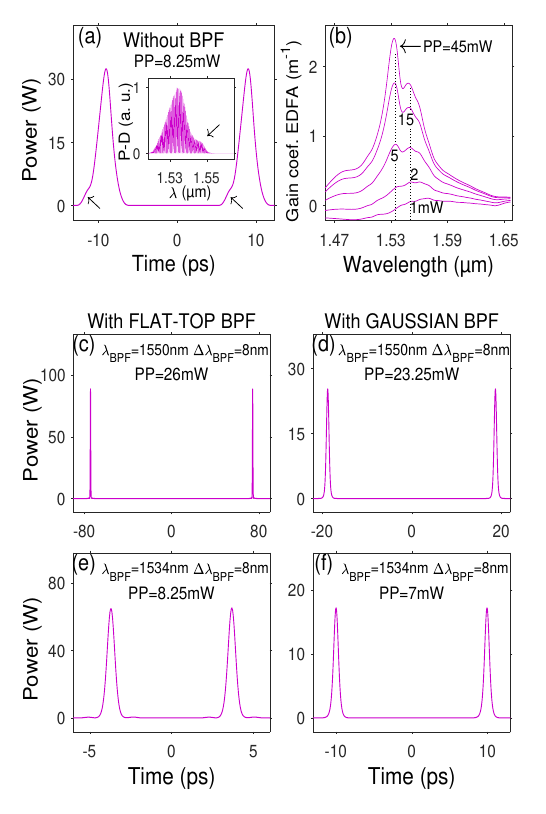}
\vspace*{0.3cm}
\caption{Intensity profiles just after pulse fragmentation, recorded at the output of the OC2,
in the region where $\Delta\lambda_{BPF} < \Delta\lambda_{c} $. (a): Pulse temporal profile for the BPF-free cavity. The inset shows the spectral power density (P-D). (b) : Gain coefficient of the EDFA. (c), (d), (e), and (f)  : Temporal profiles in presence of BPF with $\Delta\lambda_{BPF}$=8nm.}
\label{fig_pulse_above_frag_ol23_av_bif}
\vspace*{-0.2cm}
\end{figure}

We interpret the drastic decrease in FPP that we obtain by shifting the BPF to 1.534$\mu$m as follows.  Let us examine the pulses after fragmentation in the BPF-free cavity.  Their temporal profile  is shown in a wide plane in panel (a) of Fig. \ref{fig_pulse_above_frag_ol23_av_bif}, while their spectral profile is shown in the insert in that panel.
The pulse spectrum is centered at 1.534$\mu$m,  which originates from the filtering effect induced by the gain curve of the active fiber, as illustrated by the EDFA gain coefficients represented in panel \ref{fig_pulse_above_frag_ol23_av_bif} (b), obtained from the following formula for light amplification in the 1.5$\mu$m wavelength region with pumping around 980 nm \cite{Pedersen1991,Desurvire1995,Ghatak98}:
\begin{eqnarray}
g(z,\lambda)=\xi(\lambda)   N_{dop} \left[\sigma_{21}(\lambda) n_2(z)- 
\sigma_{12}(\lambda) n_1(z)  \right],
\label{gain_edfa}
\end{eqnarray}
 where $n_{1}$ and $n_{2}$ are the normalized population densities of erbium ions in levels 1 and 2 of laser transition, $\sigma_{21}$  and $\sigma_{12}$ are the absorption and emission cross-sections for a typical EDF, $N_{dop}$ is the doping concentration of erbium ions in the fiber core,  z is the longitudinal position in the fiber, and $\xi(\lambda)$ is an overlap factor taking into account that part of the light propagates outside the core where there is no erbium ion. 
As is known in the literature, the spectral gain curves of EDFA have profiles that are both very bumpy and highly dependent on PP \cite{Desurvire1995}. 
At low PP, the curve exhibits a peak gain at 1.55$\mu$m and an absorption in the vicinity of 1.53$\mu$m. However, as the PP increases, the gain curve deforms and the gain coefficient increases more in the vicinity of 1.53$\mu$m than in other regions, so that from a certain PP (which depends on the geometric parameters of the EDF and the density of erbium ions in the EDF) the peak of the gain curve shifts towards 1.53$\mu$m. 
With the set of parameters chosen to model our EDFA, this PP is around 4.5mW, which is significantly lower than the PP needed to reach the fragmentation point in our cavity. 
This is why  the peak of the EDFA gain coefficient occurs at 1.534$\mu$m during fragmentation phenomena in our modeling. 
As a result, when a 1.55$\mu$m-centered BPF is inserted into the cavity, the EDFA operates outside the bandwidth of the system (which is imposed by the BPF). 
At each cavity round trip, the amplification produced by the EDFA is then greatly reduced by the BPF. This explains why a  higher PP  is required to reach the fragmentation point when $\lambda_{BPF}=1.55\mu m$, as illustrated even more clearly by the results of pannels \ref{fig_pulse_above_frag_ol23_av_bif} (c) and \ref{fig_pulse_above_frag_ol23_av_bif} (d), obtained for $\Delta\lambda_{BPF}=8nm$.  
In contrast, by inserting a 1.534$\mu$m-centered BPF into the cavity, the EDFA operates exactly within the system bandwidth, which reduces the FPP to a level equivalent to that of the BPF-free cavity, as illustrated by the results of panels \ref{fig_pulse_above_frag_ol23_av_bif} (e) and
\ref{fig_pulse_above_frag_ol23_av_bif} (f).

\begin{figure}[h]
\centering
\includegraphics[width=.5\linewidth]{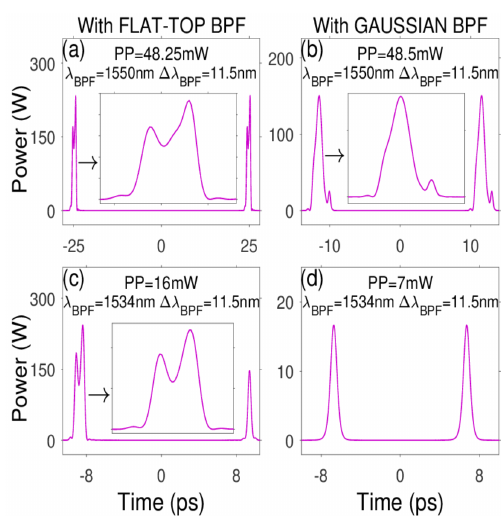}
\vspace*{-0.3cm}
\caption{Intensity profiles just after pulse fragmentation, for
$\Delta\lambda_{BPF}=11.5nm$.
 }
\label{fig_pulse_above_frag_ol23_ap_bif}
\end{figure}

Then, the question is that of the real interest of the BPF. Upon careful examination of the results of panel \ref{fig_pulse_above_frag_ol23_av_bif} (a), we note distortions (indicated by small arrows) in the pulse profiles generated in the BPF-free cavity, which we attribute to the roughness of the EDFA gain curve.  In contrast, panels \ref{fig_pulse_above_frag_ol23_av_bif} (e) and \ref{fig_pulse_above_frag_ol23_av_bif} (f) show that the presence of a BPF removes these profile distortions.  However, this is only possible for bandwidth values below the critical point. Indeed, for $\Delta\lambda_{BPF} > \Delta\lambda_{c} $,
the pulse profiles are distorted so severely that the action of the BPF alone becomes 
insufficient to restore a profile of good quality, as illustrated by the results of panels (a), (b), and (c) in Fig. \ref{fig_pulse_above_frag_ol23_ap_bif}.

Pulse distortion was pointed out in previous work, with a two-dimensional optimization technique (2D-OT) proposed to avoid it \cite{TchofoDinda2023}.
The key element of this technique is a tunable bandwidth BPF, as this component provides an additional parameter that allows the system configuration to be adjusted using two parameters: the PP of the laser and the bandwidth of the BPF.
The 2D-OT is illustrated in panels \ref{fig_pf_vs_del_ol_2023} (a) and \ref{fig_pf_vs_del_ol_2023} (b) using line segments bearing vertical and horizontal arrows, which schematically indicate the procedure for configuring the cavity with a BPF with 
$\Delta\lambda_{BPF} =11.5nm$, which is the same bandwidth as that of Fig. \ref{fig_pulse_above_frag_ol23_ap_bif} obtained by the conventional fragmentation technique. The 2D-OT takes place in two stages.

(i) First, the BPF bandwidth is adjusted to a value $\Delta\lambda_{opt}$ lower than the
 critical point 
(i.e., $\Delta\lambda_{BPF} = \Delta\lambda_{opt}=8nm$) and the PP is adjusted to the mode-locking threshold to generate a single pulse in the cavity, which corresponds to the configuration labeled ‘I’ in panels \ref{fig_pf_vs_del_ol_2023} (a) and  \ref{fig_pf_vs_del_ol_2023} (b).
Then, as in the conventional technique, the PP is gradually raised until the pulse is 
fragmented. The first step then ends with a cavity lying in an intermediate configuration labeled $\Pi$ in panels \ref{fig_pf_vs_del_ol_2023} (a) and \ref{fig_pf_vs_del_ol_2023} (b), which is in the multi-pulse regime but with a bandwidth that is well below the critical point and below the desired bandwidth.

(ii) In the second step, the PP is maintained at its final value of the first step, and the filter bandwidth is increased gradually until reaching the desired value, 11.5nm, which corresponds to the configuration labeled F in panels \ref{fig_pf_vs_del_ol_2023} (a) and \ref{fig_pf_vs_del_ol_2023} (b).

\begin{figure}[h]
\centering
\includegraphics[width=.55\linewidth]{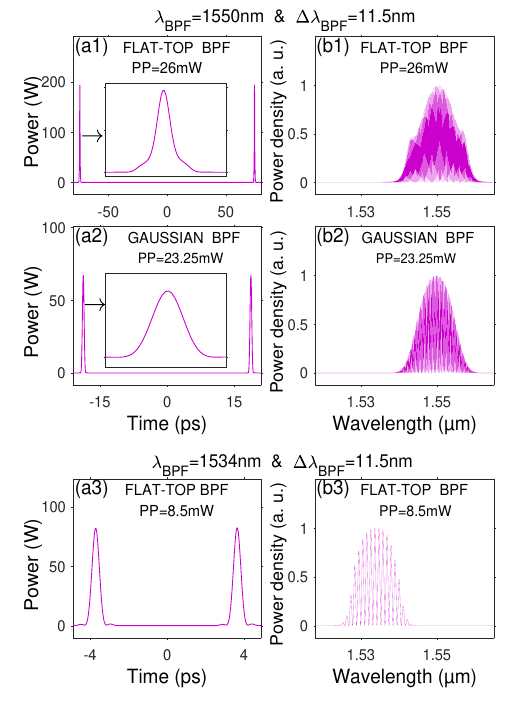}
\vspace*{0.1cm}
\caption{Intensity profiles obtained using the 2D-OT.}
\label{fig_pulse_above_frag_ol23_ap_bif_2D}
\end{figure}

The 2D-OT results for three cavity configurations considered above, namely, [FLAT-TOP BPF, $\Delta\lambda_{BPF} = 11.5nm $, $\lambda_{BPF} = 1.55\mu m $], [Gaussian BPF, $\Delta\lambda_{BPF} = 11.5nm $, $\lambda_{BPF} = 1.55\mu m $],  and [FLAT-TOP BPF, $\Delta\lambda_{BPF} = 11.5nm $, $\lambda_{BPF} = 1.534\mu m $], are shown in Fig. \ref{fig_pulse_above_frag_ol23_ap_bif_2D}.
 The comparison between the results of the panels \ref{fig_pulse_above_frag_ol23_ap_bif_2D} (a1) and \ref{fig_pulse_above_frag_ol23_ap_bif} (a), \ref{fig_pulse_above_frag_ol23_ap_bif_2D} (a2) and \ref{fig_pulse_above_frag_ol23_ap_bif} (b), and \ref{fig_pulse_above_frag_ol23_ap_bif_2D} (a3) and \ref{fig_pulse_above_frag_ol23_ap_bif} (c), shows that distortions have been significantly removed in the intensity profiles generated by 2D-OT, for a bandwidth above the critical point.
But, previous work has only provided the theoretical demonstration of 2D-OT from numerical simulations \cite{TchofoDinda2023}.  In the following, we present the experimental demonstration of the 2D-OT.

\section{Experimental results}

\begin{figure}[!h]
\centering
\includegraphics[width=.72\linewidth]{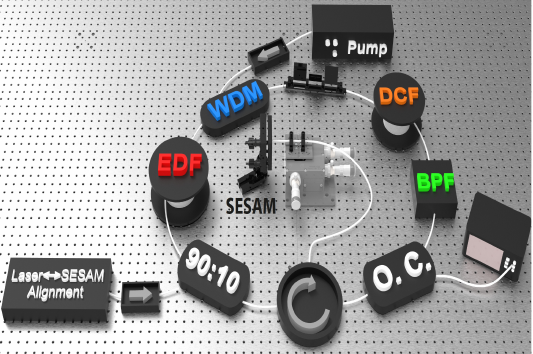}
\caption{Experimental setup of the laser cavity. 
  Intensity profile and spectra are respectively measured with an optical spectrum analyser and a 63 GHz oscilloscope.
}
\label{exp_setup}
\end{figure}

Figure \ref{exp_setup} shows the diagram of the experimental device which we have constructed, inspired by the theoretical model of Fig. \ref{schema_cavite_lineaire}. 

\begin{figure}[htb]
\centering
\includegraphics[width=.72\linewidth]{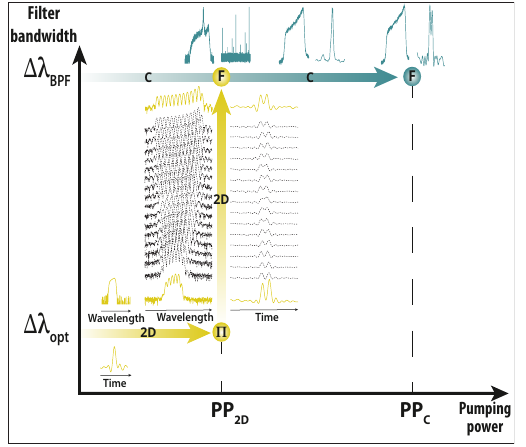}
\caption{
Experimental results showing access to multi-pulse regime by the 
conventional procedure (path labeled "C") and the 2D-OT (path labeled "2D"). $\Delta\lambda_{opt}$=2.4nm.  $\Delta\lambda_{BPF}$=6.4nm.
 PP$_{2D}$=179.3mW. PP$_{C}$=338mW.
 }
\label{Schema_methode_2D_V2_new}
\vspace*{-0.3cm}
\end{figure}

In Fig. \ref{exp_setup}, a 1.2 meter EDF pumped with a 980nm laser diode is used for amplification. A SESAM (SEmiconductor Saturable Absorber Mirror) with a relaxation time of 1ps is used with a butt-coupling technique to perform the mode locking. An external continuous laser (Laser $\leftrightarrow$ SESAM alignment) is coupled with a 90:10 coupler to perform and check the alignment of the SESAM. This laser is only used for alignment, and is switched off for all subsequent operations in the cavity. A 20\% optical coupler (O.C.) is placed at the input of the BPF to characterize the pulse profile before it passes through the BPF. The spectral profile is measured using an optical spectrum analyser, while the temporal profile is acquired using a 63 GHz oscilloscope. The latter is not able to completely resolve the time profile of the pulses, but its resolution is sufficient to detect if the multi-pulse regime is reached. The spectral filtering is performed with a commercial BPF (from Nano-Giga Inc.), with a tunable center wavelength between 1500nm and 1600nm, and a tunable bandwidth between 1nm and 100nm.
We set the BPF center wavelength to 1.55 $\mu$m, which corresponds to the operating wavelength of our SESAM, and the BPF bandwidth to 6.4nm, so that it is around or above the experimental critical point of our cavity, $\Delta\lambda_{c}$.
For the conventional method, we gradually increase the PP until fragmentation occurs, at point F 
located at the end of the path (labeled "C") in Fig. \ref{Schema_methode_2D_V2_new}, which corresponds to a FPP of 338mW.

\begin{figure}[htb]
\centering
\includegraphics[width=.65\linewidth]{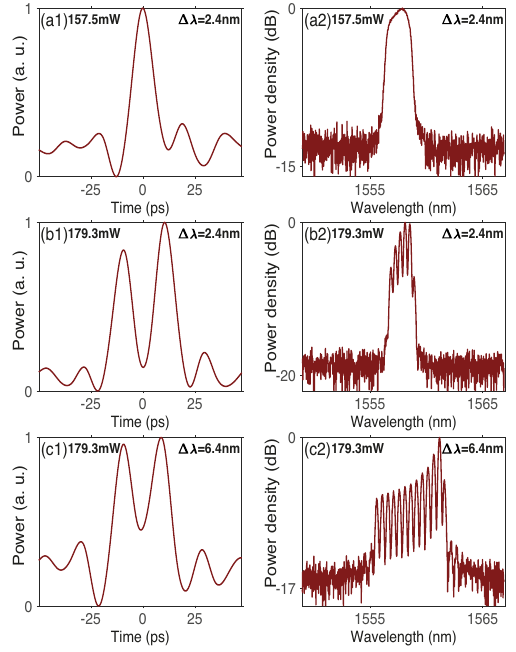}
\vspace*{-0.2cm}
\caption{Temporal and spectral profiles of the structures generated at the three main steps of the 2D procedure illustrated in Fig. \ref{Schema_methode_2D_V2_new}.
(a1)-(a2): structure obtained just above the mode-locking threshold.
(b1)-(b2): structure obtained at the intermediate point ($\Pi$).
(c1)-(c2): structure obtained at the end point (F).
 }
\label{Methode_2D_2x2_V2_new}
\end{figure}

For the 2D-OT, the filter bandwidth is first set at a value sufficiently lower than 6.4nm, namely  $\Delta\lambda_{opt}$=2.5nm, 
which can significantly lower the FPP. Then, then PP is increased to reach the fragmentation point (point $\Pi$ in the yellow path in Fig. \ref{Schema_methode_2D_V2_new}), after which the bandwidth is gradually increased up to $\Delta\lambda_{BPF}$=6.4nm (point F, at the end yellow path in Fig. \ref{Schema_methode_2D_V2_new}). The FPP in this 2D-OT is 179.3 mW, which is almost half of the FPP in the conventional procedure.
Figure \ref{Methode_2D_2x2_V2_new} clearly shows the intensity profiles corresponding to the main steps of the 2D procedure shown in Fig. \ref{Schema_methode_2D_V2_new}.

\section{Conclusion}
To conclude, we have shown that a doubly tunable BPF, i.e. with adjustable center wavelength and bandwidth, opens up new routes to the generation of multi-pulse structures at lower pump power. 
The first optimization consists in carefully adjusting the central wavelength of the filter 
with respect to the peak of the gain curve of the active fiber.
Another optimization takes places in a two-dimensional parameter space involving the laser 
pumping power and the filter bandwidth. We experimentally confirm that the latter reduces 
the required pump power by one half, compared to that required by the conventional
 operation based solely on a pump power increase.
Such 2D approach should facilitate the access and manipulation of multiple 
dissipative solitons regimes in ultrafast fiber lasers. 

\vspace*{0.3cm}

Funding : Agence Nationale de la Recherche (ANR-15- IDEX-
0003, ANR-17-EURE-0002); iXCore Research Foundation; Conseil
régional de Bourgogne; FEDER (fonds européen de développement régional).


Disclosures : The authors declare no conflicts of interest.

\printbibliography

\end{document}